\documentclass[letterpaper,onecolumn,11pt]{quantumarticle} % use larger type; default would be 10pt
\pdfoutput=1

\usepackage[utf8]{inputenc} % set input encoding (not needed with XeLaTeX)
\usepackage[T1]{fontenc} 
\usepackage{amsmath}
\usepackage{hyperref}
\hypersetup{
	pdfpagemode=UseOutlines,
	bookmarksnumbered=true
}

\usepackage{graphicx} % support the \includegraphics command and options
\usepackage{subcaption}

\usepackage{booktabs} % for much better looking tables
\usepackage{array} % for better arrays (eg matrices) in maths
\usepackage{paralist} % very flexible & customisable lists (eg. enumerate/itemize, etc.)
\usepackage{verbatim} % adds environment for commenting out blocks of text & for better verbatim
\usepackage[ruled,vlined,linesnumbered]{algorithm2e}
\usepackage{amssymb}

\usepackage{tikz}
\usepackage{pgfplots}

\newcommand{\logicalerror}{P(p)}
\newcommand{\physicalerror}{p}
\newcommand{\minfailwt}{w_\text{min}}

\newcommand{\failingset}{\mathcal{F}}
\newcommand{\markovgraph}{\mathcal{G}}

\begin{document}

\title{Improved Methods for Determining Quantum Error Correcting Code Performance and Fault Tolerance}

\author{Michael Mullan}
%\affiliation{Northrop Grumman}
\email{Michael.Mullan@ngc.com}

\author{Matthew Weippert}
%\affiliation{Northrop Grumman}

\author{Winton Brown}

\author{The Northrop Grumman Quantum Computing Team}
\affiliation{Northrop Grumman Systems Corporation, Aurora, CO}

\maketitle

\begin{abstract}

One of the central challenges in quantum error correction is determining the performance of a code in the low-error regimes needed to implement utility-scale computations.  While performance at these error rates is not amenable to direct Monte Carlo simulation, it can be extrapolated from simulations at higher logical error rates, assuming the logical error rate scales predictably with increasing distance or decreasing physical error rate. However, the expected scaling depends sensitively on the minimum weight of uncorrectable error patterns. In many cases, the minimum weight is unknown since it depends not only on the theoretical code distance, but also on details of the implementation. Markov chain Monte Carlo (MCMC) methods, as adapted to quantum error correction by Bravyi and Vargo \cite{bravyi2013simulation}, provide a way to estimate logical failure rates in these low-error regimes via simulation. While offering significant gains over Monte Carlo, the described Metropolis algorithm makes small changes to the current logical failure patterns which results in slow convergence. In this paper, we argue that typical failure patterns include a large number of easily correctable errors that coexist alongside a malignant core. This observation motivates two new approaches to better evaluate code performance.  First, we describe a pruning algorithm designed to obviate these correctable errors and focus on the problematic low-weight core. Second, we develop a novel family of Metropolis-Hastings algorithms, referred to as subregion MCMC.  This technique is parameterized by the fraction of the error pattern that is resampled at each step, effectively interpolating between Monte Carlo and single step MCMC.  We show that a judicious choice of this parameter results in far faster convergence than prior work.   
\end{abstract}

%%%%%%%%%%%%%%%%%%%%%%%%%%%%%%%%%%%%%%%%%%%%%%%%%%%%%%%%%%%%%%%%%%%%%%%%%%%%%%%%%%%%
\section{Introduction}
%%%%%%%%%%%%%%%%%%%%%%%%%%%%%%%%%%%%%%%%%%%%%%%%%%%%%%%%%%%%%%%%%%%%%%%%%%%%%%%%%%%%

Estimates of the number of operations needed to execute the first utility-scale quantum algorithms often exceed $10^{10}$ \cite{lee2021even, gidney2025factor} necessitating logical error rates below $10^{-10}$.  Most simulations of quantum error correcting codes
use Monte Carlo methods, where each run of a circuit is subject to an independent, random distribution of faults chosen according to an approximation of the underlying physical error model.  While these methods are straightforward to implement and parallelize, achieving a confidence of 10\% in a logical error rate of $10^{-10}$ requires approximately $10^{12}$ runs, which can quickly become unwieldy for complex codes and infeasible for circuits containing non-Clifford gates.  

One common solution is to determine a code's logical error rate at several higher physical error rates (and/or lower distances), and extrapolate the performance to lower error rates using the expected asymptotic scaling, 
\begin{equation} \label{eq:scaling}
\logicalerror = \alpha \physicalerror^{\minfailwt}
\end{equation}
where $\logicalerror$ is the logical error rate, $\physicalerror$ is the physical error rate, $\alpha$ is a constant that depends on the code and implementation details, and $\minfailwt$ is the minimum weight over all uncorrectable error patterns, i.e., error patterns that result in a logical failure.  Note that $\minfailwt \le t+1$, where $t = \left\lfloor (d-1)/2 \right\rfloor$ and $d$ is the distance of the code.  In a particular instantiation of a code, $\minfailwt$ depends not only on $d$, but also on the syndrome extraction circuit as well as the decoding algorithm; instantiations where $\minfailwt = t+1$ are called fault-tolerant. In a simulation, an unexpectedly low $\minfailwt$ could come from numerous sources including problems with code design, a syndrome extraction circuit that spreads errors, a decoder that doesn't correct up to the full distance, or simply a bug in the code's implementation. Unfortunately, these issues do not always clearly manifest in simulations at higher error probabilities, but instead appear as subtle deviations to the expected scaling, which may be difficult or impossible to distinguish from Monte Carlo statistical noise.

Predicting logical error rates in the low probability regime then requires either A) methods to determine $\minfailwt$ so that the scaling relationship (Eq. \eqref{eq:scaling}) can be confidently applied and/or B) better simulation techniques designed for exploring this regime directly. 

Techniques such as direct counting \cite{aliferis2005quantum} and subset sampling \cite{heussen2024dynamical} can be useful for particular codes but are not broadly applicable.  Markov chain Monte Carlo (MCMC) \cite{bravyi2013simulation} has emerged as a promising solution, as it's designed for sampling rare events from arbitrary probability distributions.  The standard implementation of MCMC in quantum error correction uses the Metropolis algorithm \cite{metropolis1953equation, hastings1970monte} to draw samples;  however, as we'll discuss, this algorithm is not efficient, as it takes small steps and accepts newly generated samples with low probability.  Increasing both the step size and acceptance rate is highly desirable in a number of fields and has been a topic of ongoing research.  Many such advances have been targeted at specific types of distributions or specific physical systems, e.g. Gibbs sampling \cite{geman1984stochastic}, slice sampling \cite{neal2003slice}, and Hamiltonian Monte Carlo \cite{duane1987hybrid}.  In the context of quantum error correction specifically, Refs. \cite{mayer2025rare, beverland2025fail} have adapted MCMC to the circuit model and improved its applicability to a variety of codes.  

Through two complementary techniques, this paper shows both how to use simulations at higher error rates to estimate $\minfailwt$, as well as how to more efficiently simulate at lower error rates, via a significant improvement to the speed of Markov chain Monte Carlo.   Our methods are inspired by one key observation about uncorrectable error patterns in many codes: At physical error rates not too far from threshold, in the regime in which we expect near and intermediate term machines to operate, an uncorrectable error set contains both a small, malignant core, as well as a large number of easily correctable errors, which we'll refer to as ``fluff''.  We'll describe ways to cut through this fluff and focus on the core, malignant pattern.  Indeed, the manipulation of an existing, known failure pattern will prove to be very powerful.  

Ultimately this paper is about ensuring that a code, when implemented in practice, hits its predicted performance targets.  As quantum computing increasingly moves out of small laboratory experiments and into the realm of engineering, we expect concerns like these, and the ability to accurately and swiftly evaluate the performance of arbitrary codes, to be increasingly important.  

All techniques in this paper are simulated in the full circuit model using QVM (Quantum Virtual Machine), Northrop Grumman's high-performance quantum circuit simulation software.  We associate an error probability with each circuit location, which corresponds to an individual single- or two-qubit gate, or an identity gate where no other gate is applied.  We use a depolarizing error model with the same error probability for each location, subdivided equally over each possible Pauli-error.  Our techniques are, however, applicable to more general error models.  We simulate both the standard, un-rotated surface code \cite{dennis2002topological}, decoded with Minimum Weight Perfect Matching (MWPM) implemented in Blossom V \cite{kolmogorov2009blossom} and a two-level concatenated Bacon-Shor code \cite{bacon2006operator} decoded with a basic lookup table. We simulate $d$ rounds of the appropriate syndrome extraction circuit, followed by a perfect, error-free measurement.

%%%%%%%%%%%%%%%%%%%%%%%%%%%%%%%%%%%%%%%%%%%%%%%%%%%%%%%%%%%%%%%%%%%%%%%%%%%%%%%%%%%%
\section{Pruning Method} 
\label{sec:pruning}
%%%%%%%%%%%%%%%%%%%%%%%%%%%%%%%%%%%%%%%%%%%%%%%%%%%%%%%%%%%%%%%%%%%%%%%%%%%%%%%%%%%%

In this section, we start by examining the typical characteristics of failure patterns that occur in error rate regimes of interest.  Our observations will suggest a straightforward way to use a series of high-$p$ Monte Carlo simulations to demonstrate that an implementation of an error correcting code achieves the expected $\minfailwt$, allowing us to extrapolate to low-$p$ performance. In the next section, we'll use these same observations to improve the performance of MCMC significantly.  

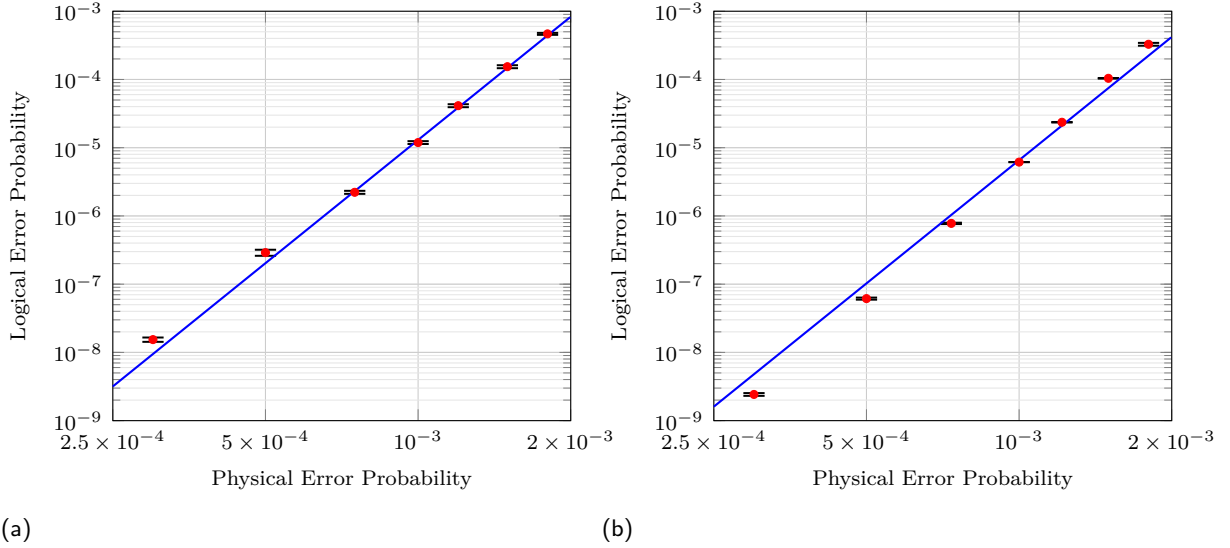
\begin{figure}
\centering
	\begin{subfigure}{0.49\textwidth}
		\centering
		\begin{tikzpicture}
\begin{loglogaxis}[
    width=\linewidth,
    height=7cm,
    xlabel={Physical Error Probability},
    ylabel={Logical Error Probability},
    grid=both,
    minor grid style={gray!20},
    major grid style={gray!40},
    xtick={2.5e-4,5e-4,1e-3,2e-3},
    xticklabels={{$2.5\times10^{-4}$},{$5\times10^{-4}$},{$10^{-3}$},{$2\times10^{-3}$}},
    minor x tick num=8,
  xmin=2.5e-4,
    xmax=2e-3,
ymin=1e-9,
ymax=1e-3,
    tick label style={font=\scriptsize},
   xlabel style={font=\scriptsize},
  ylabel style={font=\scriptsize}
]

\addplot[
    only marks,
   color=red,
    mark=*,
    mark size=1.5pt,          % smaller points
    % --- error bar configuration ---
    error bars/.cd,
    y dir=both,
    y explicit,
    error bar style={color=black, line width=0.75pt},
    error mark=|,
    error mark options={
        line width=0.75pt,
        mark size=4pt        % length of the horizontal caps
    },
]
table[x=x, y=y, y error=dy] {
x       y          dy
1.8e-3  4.66e-4    1.5378e-5
1.5e-3  1.54e-4    7.70e-6
1.2e-3  4.13e-5    2.065e-6
1.0e-3  1.19e-5    5.712e-7
7.5e-4  2.22e-6    1.1322e-7
5.0e-4  2.90e-7    2.958e-8
3.0e-4  1.54e-8    1.1242e-9
};

\addplot[
    domain=2.5e-4:2e-3,
    samples=200,
    smooth,
    thick,
    color=blue,
]
{1.29923e13 * x^6};

\end{loglogaxis}
\end{tikzpicture}
		\caption{}
	\end{subfigure}
	\hfill
	\begin{subfigure}{.49\textwidth}
		\centering
		\begin{tikzpicture}
\begin{loglogaxis}[
    width=\linewidth,
    height=7cm,
    xlabel={Physical Error Probability},
    ylabel={Logical Error Probability},
    grid=both,
    minor grid style={gray!20},
    major grid style={gray!40},
    xtick={2.5e-4,5e-4,1e-3,2e-3},
    xticklabels={{$2.5\times10^{-4}$},{$5\times10^{-4}$},{$10^{-3}$},{$2\times10^{-3}$}},
    minor x tick num=8,
    xmin=2.5e-4,
    xmax=2e-3,
ymin=1e-9,
ymax=1e-3,
    tick label style={font=\scriptsize},
  xlabel style={font=\scriptsize},
  ylabel style={font=\scriptsize}
]

% Data + error bars (second dataset)
\addplot[
    only marks,
    mark=*,
    mark size=1.5pt,
    color=red,
    mark options={draw=red, fill=red},
    % --- error bar configuration ---
    error bars/.cd,
    y dir=both,
    y explicit,
    error bar style={color=black, line width=0.75pt},
    error mark=|,
    error mark options={
        line width=0.75pt,
        mark size=4pt
    },
]
table[x=x, y=y, y error=dy] {
x           y              dy
1.8e-3      3.28e-4        1.5744e-5     
1.5e-3      1.04e-4        1.04e-6       
1.21476e-3  2.36e-5        2.124e-7      
1.0e-3      6.15e-6        1.1685e-8     
7.35042e-4  7.78e-7        1.80596e-8    
5.0e-4      6.143e-8       2.15005e-9    
3.0e-4      2.42e-9        1.13256e-10   
};

% Same power-law line as before
\addplot[
    domain=2.5e-4:2e-3,
    samples=200,
    smooth,
    thick,
    color=blue,
]
{6.56765e12 * x^6};

\end{loglogaxis}
\end{tikzpicture}
		\caption{}
	\end{subfigure}
\caption{A $d = 11$ surface code decoded using MWPM.  In (a) the decoder fails to take into account one of the hooks.  The first five highest $p$ points are fit to Eq. \eqref{eq:scaling} with $w_{min} = 6$, which predicts a logical error rate of $\logicalerror =9.47 \times 10^{-9} $ at $\physicalerror = 3 \times 10^{-4}$; however the Monte Carlo simulation predicts $\logicalerror=1.54 \times 10^{-8}$.  Notice that, until points six and seven are included, this fit looks perfectly fault tolerant, and it is not obvious that there is a bug somewhere in our implementation.  In (b), we correct the hook and the fit (still based only on the first five points) now predicts a logical error rate of $\logicalerror = 4.79 \times 10^{-9}$.  Black error bars represent one standard error.}
\label{fig:surface_hook}
\end{figure}

Consider an example where extrapolation via Eq. \eqref{eq:scaling} from high $p$ to low $p$ is problematic. Figure~\ref{fig:surface_hook}(a) is a plot of a distance 11 surface code, where we have intentionally introduced non-fault tolerance into the decoder by improperly accounting for one of the hooks. (see Ref. \cite{dennis2002topological} for a discussion of hooks generally).  Here we fit Eq. \eqref{eq:scaling} to the first five data points, i.e., those simulated at the five highest physical error rates.  The line fits these five points quite well, and in the absence of any additional data, we'd take this as good evidence that the code is performing fault-tolerantly.   However, notice that the next two data points, both simulated at lower $p$, fall off the line, and we can now see that our fit is underestimating logical error rates.  This is a significant issue in practice as it will result in system designs with undersized codes that fail to execute the targeted algorithm with a sufficient success rate.  In contrast, in figure~\ref{fig:surface_hook}(b), we correctly account for the missing hook, and unsurprisingly, see a significant difference in logical error rates.  In this case, the low probability points outperform the fit's prediction.

This pattern - where non-fault-tolerant behavior is not evident at high probabilities - is unfortunately quite common.  Subtle non-fault tolerance, where only a few select low-weight error patterns cause logical errors, is more likely to affect the apparent scaling at low physical error rates.  Indeed, if we assume, for simplicity, that each circuit location has a single failure mode and a uniform probability of error, then the logical error rate is given by 
\begin{equation*}
\logicalerror = \sum_w f(w) \binom{N}{w} p^w (1-p)^{N-w}
\end{equation*}
where $f(w)$ is the fraction of weight-$w$ error patterns that are uncorrectable, and $N$ is the number of circuit locations.  At high $p$, this sum will contain significant contributions from terms where $w > t$, especially when $f(w)$ is large.  On the other hand, at low $p$, these terms will become increasingly negligible and the sum will be dominated by the term with $w = \minfailwt$, effectively reducing to Eq. \eqref{eq:scaling}.

This makes alternate methods of searching for non-fault-tolerant error patterns highly desirable. Searching for low-weight patterns via Monte Carlo is infeasible, especially in the full circuit model, since the typical number errors applied at commonly considered physical error rates is often very high. For example, our implementation of the distance-11 surface code with $d = 11$ rounds of syndrome extraction contains 26,694 locations.  Thus, at a physical error rate of $p=10^{-3}$ we expect roughly 27 errors, which is about four standard deviations away from weight $t=5$ error patterns.    In the the two-level concatenated Bacon-Shor code, the situation is worse.  In the 5-on-5 code, for example, there are 204,750 locations and so we expect roughly 205 errors, but the code can fail with only 9. Indeed, for codes that are not especially small, we \textit{never} expect to see less than $t+1$ errors, even if the code performs poorly. This is illustrated in more detail in table~\ref{tab:counts}.

\begin{table}[h] \label{tab:codes}
	\centering
	\begin{tabular}{|l|r|r|r|r|}
		\hline
		Code &  $\minfailwt$ & Num. Locations & Exp. Num. Errors & Prob. $< \minfailwt$ \\
		\hline
		Surface(5) & 3 & 2232 & $2.2 \pm 1.5$ & $< 10^{0}$ \\
		Surface(7)  & 4 & 6514 & $6.5 \pm 2.6$ & $< 10^{0}$ \\
		Surface(9)  & 5 & 14316 & $14.3 \pm 3.8$ & $< 10^{-2}$ \\
		Surface(11) & 6 & 26694 & $26.7 \pm 5.2$ & $< 10^{-6}$ \\
		Surface(13) & 7 & 44704 & $44.7 \pm 6.7$ & $< 10^{-12}$ \\
		Concatenated BS(3 on 3) & 4 & 27439 & $27.4 \pm 5.2$ & $< 10^{-8}$ \\
		Concatenated BS(5 on 5) & 9 & 204750 & $204.8 \pm 14.3$ & $< 10^{-75}$ \\
		Concatenated BS(7 on 7) & 16 & 798161 & $798.2 \pm 28.2$ & $< 10^{-315}$ \\
		\hline
	\end{tabular}
	\caption{An accounting of the minimum failure weight, number of circuit locations, the expected number of physical errors, and the probability of obtaining less than $\minfailwt$ errors during $d$ rounds of syndrome extraction at $p = 10^{-3}$ for the codes studied here.  Notice that as code size grows, the odds of identifying non-fault-tolerant behavior based on error counting alone rapidly declines.  For the Bacon-Shor codes, $\minfailwt < t + 1$ because the decoder used operates at each level of concatenation independently.}  
	\label{tab:counts}
\end{table}

\begin{figure}  
\centering
\includegraphics[width =\linewidth]{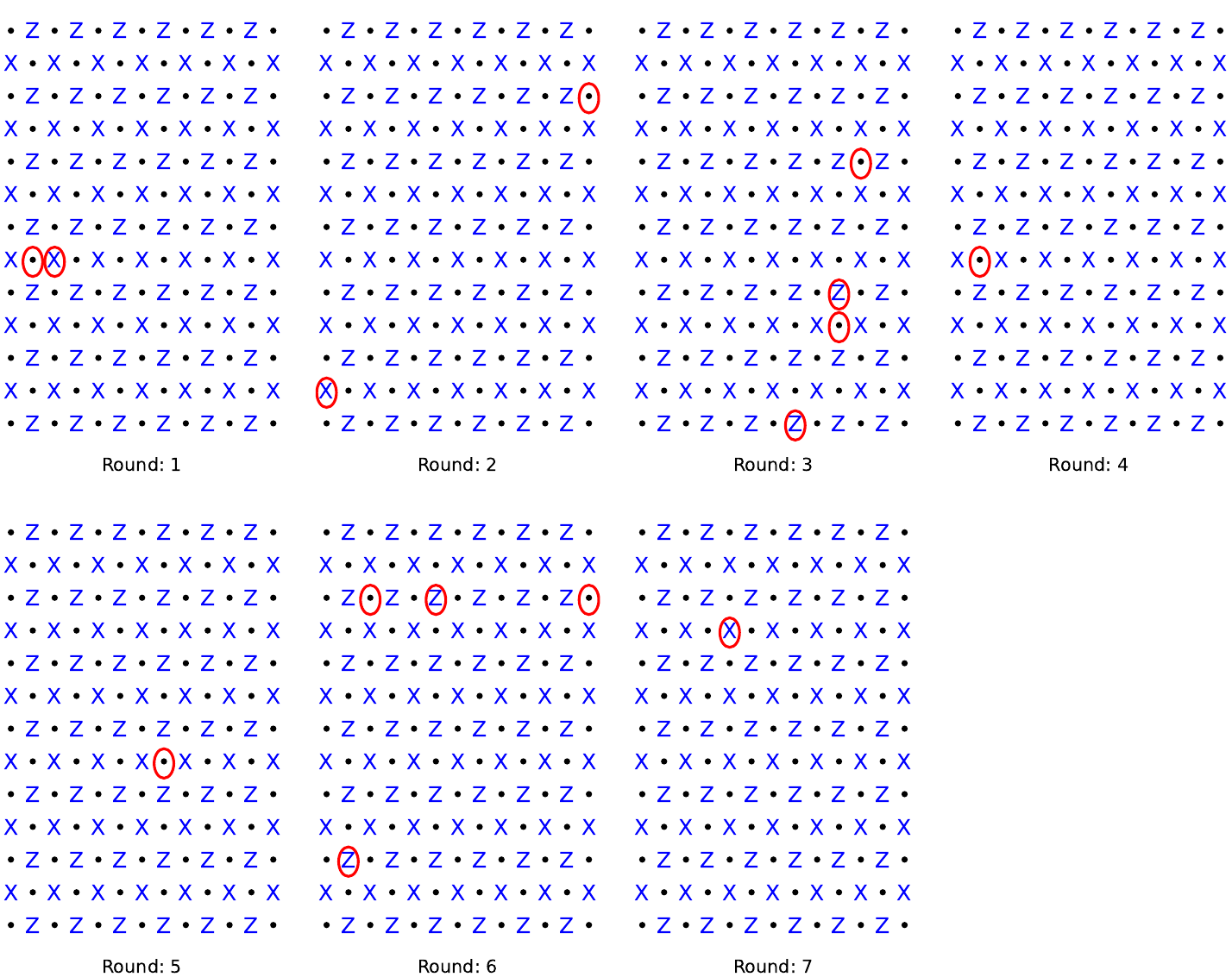}
\caption{A randomly chosen failure pattern on a $d = 7$ surface code.  Red circles indicate where an error was first generated; these may subsequently propagate via gates or the decoder.  Note that the ratio of fluff errors to malignant core errors is worse at higher distances and physical error rates.}
\label{fig:surface_errors}
\end{figure}

Consider, however, figure~\ref{fig:surface_errors}, which illustrates a typical failure pattern on a $d = 7$ surface code patch at $p=10^{-3}$.  Notice that the vast majority of errors are isolated and easily corrected with MWPM.  This behavior is precisely what is expected when errors are distributed uniformly at random over what is essentially a three dimensional volume, and failure modes consist of one dimensional lines.  These scattered, easily correctable errors are what we refer to as fluff.

Because of these considerations, rather than searching for low-weight failing patterns, we focus on testing if a given, high-weight failing pattern should be correctable. Our strategy is to, over a series of rounds, trim away potential fluff errors and examine the errors that remain. There is no clear analytical way of determining which errors are fluff and which are part of the uncorrectable core.  Therefore, during each round, we remove a some small subset of physical errors, $S$, from the full pattern $E$, then run the error correction circuit and decoder to test for a logical error.  If there is still a logical error, all of the errors in $S$ are considered fluff; otherwise, they may form part of the core, and are added back to $E$. Specifically, we proceed according to algorithm~\ref{alg:pruning}.

\begin{algorithm} 
\caption{Prune Errors}
\label{alg:pruning}
\KwData{Circuit $C$}
	Generate $M$ failure patterns using a Monte Carlo simulation of circuit $C$\;
	\For{each failure pattern $E$} {
	\For{$K$ rounds} {
		Remove a subset $S$ of errors from $E$, so that $E = E \setminus S$\;
		Run circuit $C(E)$\;
			\If{$E$ is correctable} {
				$E = E \cup S$\;
			}
	}
	}

\end{algorithm}

There are multiple potential removal strategies that can be used in line (4).  The simplest option is to progress sequentially through $E$ and attempt to remove each error individually. While this is time-efficient, it may fail to remove interdependent error sets.  Consider, for example, three $X$ errors that happen on a single qubit which is part of the core failure pattern.  Removal of any one of these errors will avert the failure, and so, our procedure will consider that error significant.  However, removal of any two $X$ errors is acceptable, as this results in the same core pattern.  Therefore, to confidently prune such interdependent error sets, we instead remove a random number of errors at random locations during each round.  

Using this approach, we've found that, in practice, a relatively small removal size  ($<5$) and a small number of rounds (1000-10000) frequently reduces the starting error pattern to its minimum size.  Even when the resulting pattern is not minimal, algorithm~\ref{alg:pruning} dramatically shrinks its size. At this point, if additional confidence is desired, it is often possible to rerun the circuit on all subsets of errors whose size is less than the expected $\minfailwt$.

\begin{figure} 
\centering
\includegraphics[width =\linewidth]{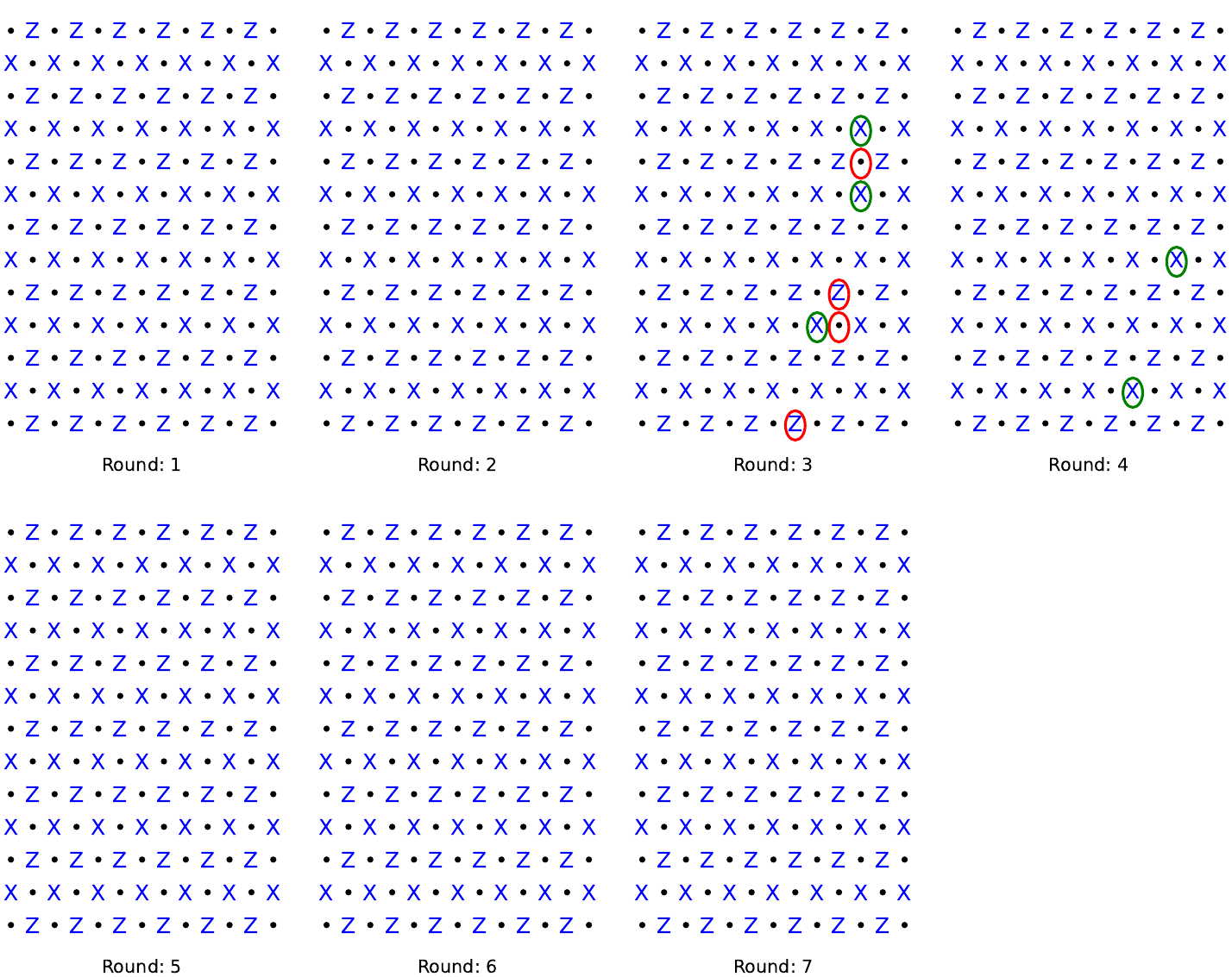}
\caption{The same error pattern shown in Fig.~\ref{fig:surface_errors} pruned via algorithm~\ref{alg:pruning}, leaving behind only the weight $t = 3$ core.  Again, red circles indicate where an error was first generated, whereas green circles correspond to the resulting error syndrome.  The two vertically adjacent circles correspond to a single two-qubit error on a CX gate.  During syndrome extraction, this error, along with the error at the bottom of the code,  spread to their neighbors on the right.  The spread of the former error results in a diagonal space-space-time hook going from (row = 10, column = 9, round = 3) to (row = 8, column = 11, round = 4).  This hook is erroneously assigned a weight of 3, causing MWPM to make the wrong correction.  This issue is much more apparent after the fluff, and consequently, the syndromes generated by the fluff, have been removed.   }
\label{fig:surface_errors_pruned}
\end{figure}

After algorithm~\ref{alg:pruning} is run to completion, if a sufficiently large number of Monte Carlo failure patterns were pruned, and none of the resulting patterns are of size $|E| \leq t$, the circuit is very likely (although not definitively) fault-tolerant.  In our experience, a few hundred such patterns typically suffices even if the pathological behavior is quite subtle. Any low-weight error pattern that is found can be run back through the circuit to determine the precise failure mechanism.  This approach, while simple, has proven to be an indispensable tool for debugging complex quantum error correcting circuits and decoding algorithms, and offers far more insight than feeding the circuit the much larger pattern containing the additional fluff.      This is illustrated in figure~\ref{fig:surface_errors_pruned} where we apply algorithm ~\ref{alg:pruning} to the error pattern in figure~\ref{fig:surface_errors}.  We see a pattern of weight $t=3$ that causes a logical failure;  by looking at the syndrome alongside the reduced failure pattern, the hook we did not properly account for becomes evident.  Note that this is the same missing hook used to generate the data in Fig.\ref{fig:surface_hook}.

As we'll discuss in section~\ref{sec:mcmc} below, algorithm~\ref{alg:pruning}  is reminiscent of the Metropolis steps that MCMC uses to sample failure patterns. Moreover, our observations here about the typical structure of error patterns will be key to increasing the efficiency of our new method.  

%%%%%%%%%%%%%%%%%%%%%%%%%%%%%%%%%%%%%%%%%%%%%%%%%%%%%%%%%%%%%%%%%%%%%%%%%%%%%%%%%%%%
\section{Subregion Markov Chain Monte Carlo}
\label{sec:subregion}
%%%%%%%%%%%%%%%%%%%%%%%%%%%%%%%%%%%%%%%%%%%%%%%%%%%%%%%%%%%%%%%%%%%%%%%%%%%%%%%%%%%%

Markov chain Monte Carlo methods combined with the splitting method enable simulations of quantum error correction circuits at low logical error rates. MCMC uses a Metropolis-Hastings algorithm to generate samples from the distribution of uncorrectable error patterns at various physical error rates. By sampling only within this space, this approach loses information about the probability of a logical failure.  However, the splitting method is then able to take samples generated from two nearby physical error rates and estimate the corresponding ratio of logical error rates. To be concrete, let $p_0,\, p_1,\,\ldots,\, p_t$ be a sequence of physical error rates such that the logical error rate, $P(p_0)$, is known. Then splitting method computes $P(p_t)$ using 
\begin{equation} \label{eq:splitting}
	P(p_t) = P(p_0) \prod_{j=1}^t \frac{P(p_j)}{P(p_{j-1})}.
\end{equation}

In section~\ref{sec:mcmc}, after describing how to construct Markov chains using a Metropolis-Hastings algorithm, we show how these samples are used to estimate the logical failure ratios in Eq. \eqref{eq:splitting}. The section concludes with a brief discussion of convergence criteria and the difficulty in estimating statistical error. 

Section~\ref{sec:mcmc} is quite general and applies to any Metropolis-Hastings approach.  As we discuss, the Metropolis-Hasting approach can be used with a variety of different proposal algorithms, each of which provide distinct Markov processes. However, once converged, each samples from the same stationary distribution over uncorrectable error patterns. In section~\ref{sec:proposal}, we describe and compare two such methods: the method described by Bravyi and Vargo \cite{bravyi2013simulation}, which we refer to as BV MCMC, and subregion MCMC, which is based on a novel family of proposal distributions. 

\subsection{Markov chains from Metropolis-Hastings} \label{sec:mcmc}

Following the notation of Ref. \cite{bravyi2013simulation}, let $\pi_j(E)$ be the probability for error pattern $E$ to occur at physical error $p_j$. Denote the set of all uncorrectable error patterns by $\failingset$, and the probability of $E$ conditioned on failure by $\pi_j(E \mid \failingset) = \pi_j(E)/\pi_j(\failingset)$, where we have used the shorthand 
\begin{equation}
	\pi_j(\failingset) \equiv \sum_{E \in \failingset}  \pi_j(E) = P(p_j).
\end{equation}
Our goal is to generate samples from the distribution of uncorrectable errors $\pi_j(E \mid \failingset)$. To this end, the Metropolis-Hastings framework provides a procedure for designing a Markov process, which we think of as a directed graph, $\markovgraph$, with vertex set $\failingset$. An edge, $E \rightarrow E'$ is in $\markovgraph$ whenever there is a non-zero transition probability, $P(E' \mid E)$, and these edges are labeled with that probability. In the context of MCMC, a Markov chain is a sequence of error patterns $E_\alpha$ obtained from a random walk on this graph.  

If the probability of transitioning from $E$ to $E'$ satisfies a ``detailed balance'' condition with respect to the desired sampling distribution,
\begin{equation} \label{eq:detailed_balance}
P(E' \mid E) \, \pi_j(E \mid \failingset) = P(E \mid E') \, \pi_j(E' \mid \failingset) \\; \ \forall E,E' \in \failingset,
\end{equation}
then $\pi_j(E \mid \failingset)$ is a stationary distribution for the Markov process.  If the Markov process is designed to satisfy detailed balance with respect to $\pi_j(E)$ it will also satisfy Eq. \eqref{eq:detailed_balance} since they differ only by an unknown normalization factor $\pi_j(\failingset) = P(p_j).$  Note that detailed balance only guarantees that $\pi_j(E \mid \failingset)$ is one of the possible stationary distributions. If the Markov process has multiple disconnected components, then the long time behavior of the random walk might depend on its starting point.  However, if the resulting Markov process is also ergodic, then it has a unique stationary distribution reachable from any point. For the finite state Markov processes described here, irreducible Markov processes are always ergodic.

The Metropolis-Hastings algorithm breaks the transition probability up into a proposal distribution $g(E \mid E')$ and an acceptance rate $A(E', E)$ so that $P(E' \mid E) = A(E', E) g(E' \mid E).$  Then Equation \eqref{eq:detailed_balance} is satisfied as long as
\begin{equation}
\frac{A(E', E)}{A(E, E')} = \frac{\pi_j(E')}{\pi_j(E)}\frac{g(E \mid E')}{g(E' \mid E)}. 
\end{equation}
Detailed balance can be enforced by using the so-called Metropolis choice for $A(E', E)$, given by
\begin{equation} 
\label{eq:acceptance}
A(E', E) = \min \left( 1, \frac{\pi_j(E')}{\pi_j(E)}\frac{g(E \mid E')}{g(E' \mid E)} \right).
\end{equation}
When generating samples in the Markov chain, the next uncorrectable error pattern depends only on the previous sample generated, $E$. The Metropolis-Hastings algorithm selects a error pattern, $E'$, sampled from $g(E' \mid E)$ which is accepted with probability $A(E', E)$. At this point, we must verify that $E'$ is uncorrectable by simulating the circuit and/or running a decoding algorithm. If $E'$ is not accepted, or if it is correctable, then the random walk returns to $E$, otherwise $E'$ is the next step.

In practice, determining when a Markov chain has sufficiently converged to its stationary distribution can be challenging. Typical recommendations include running multiple Markov chains initialized with independent samples, discarding the first half of the samples in each chain, and checking that the Gelman-Rubin diagnostic, $\hat{R}$ , is sufficiently close to 1.0 \cite{vats2021revisiting,gelman2011inference}. The Gelman-Rubin diagnostic compares the variance within each chain to the variance between all chains and decreases the more the independent chains resemble one another.  It's important to recognize that these techniques can only reliably determine lack of convergence, since the resulting random walks can get stuck in, and thoroughly sample, a small fraction of the relevant error patterns.

The logical failure ratios needed by Eq. \eqref{eq:splitting} can be computed as
\begin{equation}
	\frac{P(p_j)}{P(p_{j-1})} = \frac{\pi_j(\failingset)}{\pi_{j-1}(\failingset)} = \frac{\langle \pi_j \rangle_{j-1}}{\langle \pi_{j-1} \rangle_j},
\end{equation}
where the expected values are taken with respect to $\pi_j(E \mid \failingset)$.   That is, for a function $f: \failingset \rightarrow \mathbb{R}$,
\begin{equation} \label{eq:expected}
	\langle f \rangle_j \equiv \sum_{E \in \failingset} f(E) \pi_j(E \mid \failingset),
\end{equation}
 which can be approximated as
\begin{equation}
 \sum_{E \in \failingset} f(E) \pi_j(E \mid \failingset) \approx \frac{1}{K} \sum_{\alpha=1}^{K} f(E_\alpha),
\end{equation}
where $E_\alpha$ are samples from Markov chains at $p_j$. Note error probabilities at $p_j$ are computed for error patterns generated at $p_{j-1}$ and vice versa. A more accurate estimate for a given number of samples, $K$, was developed by Bennett \cite{bennett1976efficient}, first applied to error correction in Ref. \cite{bravyi2013simulation}, and summarized in references \cite{mayer2025rare,beverland2025fail}. Increasing the overlap between adjacent failure distributions increases the accuracy of this method; a heuristic for balancing accuracy and the number of splitting steps can be found in Ref. \cite[Eq.~(17)]{bravyi2013simulation}. 

Accurate methods for estimating statistical error when using MCMC are complicated by temporal correlations in each Markov chain \cite{bennett1976efficient}. While these correlations do not affect estimates from sufficiently long sample averages in Eq.~\eqref{eq:expected}, the variance of the resulting estimate is no longer given by $\sigma^2/K$, where $\sigma^2$ is the variance of the chain. We defer a detailed investigation of this topic for future research and use $\hat{R}$ as proxy for statistical error as well as a convergence diagnostic. This approach is supported by Ref. \cite{vats2021revisiting}, where the authors provide a direct relationship between $\hat{R}$ and the effective sample size. The effective sample size is the number of independent samples and is inversely proportional to the expected decorrelation time, or integrated autocorrelation time, of the Markov process which is roughly the expected number of steps between two states in a chain before they become effectively independent. 

\subsection{Proposal Algorithms for BV and Subregion MCMC} 
\label{sec:proposal}

BV MCMC was originally described for bit noise. In that case, new errors are generated from the previous error by selecting a single bit at random and flipping that bit. The proposal distribution is therefore $g(E' \mid E) = 1/N$. Since this is symmetric, $g(E \mid E') = g(E' \mid E)$, it drops out of the acceptance rate calculation in Eq. \eqref{eq:acceptance}. This applies directly to circuit bit flip or phase flip noise and can be adapted to circuit depolarizing noise \cite{beverland2025fail} and more general noise \cite{mayer2025rare}.

Before describing subregion MCMC, we introduce notation for a large class of circuit noise error models. Error patterns are represented by $E \in \mathbb{Z}^N$, in a circuit with $N$ locations.  Each location is thus associated with an error state, labeled by an integer,  $e_\ell $, with $e_\ell = 0$ corresponding to no error.  We consider only error models where errors at each location are independent, so that
\begin{equation}
\label{eq:newErrorProb}
	\pi_j(E) = \prod_{\ell=1}^N \pi^{(\ell)}_j(e_\ell).
\end{equation}
The error probability at each location is given by
\begin{equation}
	\pi^{(\ell)}_j(e_\ell) = \begin{cases}
		p_{j, \ell} q(e_\ell) & e_\ell > 0 \\
		(1-p_{j, \ell}) & e_\ell = 0,
	\end{cases}
\end{equation}
where $q(e_\ell)$ is the probability of picking error type $e_\ell$ conditioned on the gate at location $\ell$ failing. The splitting method requires that the gate error rate at each location be a monotonic function of the base error rate, $p_j$. For circuit depolarizing noise, $p_{j,\ell} = p_j$ and $q(e_\ell) = 1/15$ for two-qubit gates, $q(e_\ell) = 1/3$ for all single qubit gates, and $q(e_\ell) = 1$ for preparation and measurement errors. 

Our subregion method modifies the BV approach in two ways. First, we allow larger changes to the current error pattern by selecting more locations to update. Second, rather than "flipping" errors, we resample the errors at these locations, potentially with a different base error rate. Our proposal algorithm is as follows:  
\begin{enumerate}
\item Select a random subregion of the code.  Each location in the circuit is selected with probability $p_r$, the region rate.
\item Outside this region, errors in $E$ remain the same.  Inside the region, resample locations according to the error model using a base error rate of $p_f$, the flip rate.
\end{enumerate}

To calculate the acceptance rate $A(E', E)$ for Eq. \eqref{eq:acceptance}, we first look at the ratio of proposal distributions. Consider locations where the error type is the same for $E'$ and $E$. This occurs either when location $\ell$ was not selected for resampling, or when the resampled error state matches the previous state at $\ell$. On the other hand, any locations where the error types differ must have been selected for resampling. Grouping the factors by these cases we have
\begin{align}
\frac{g(E' \mid E)}{g(E \mid E')} &= \prod_{e_\ell \neq e'_\ell} \frac{p_r \pi_f^{(\ell)}(e'_\ell)}{p_r \pi_f^{(\ell)}(e_\ell)}  \prod_{e_\ell = e'_\ell} \frac{(1-p_r) + p_r \pi_f^{(\ell)}(e'_\ell)}{(1-p_r) + p_r \pi_f^{(\ell)}(e_\ell)} \\ 
  &= \prod_{e_\ell \neq e'_\ell} \frac{\pi_f^{(\ell)}(e'_\ell)}{\pi_f^{(\ell)}(e_\ell)}  \prod_{e_\ell = e'_\ell} \frac{\pi_f^{(\ell)}(e'_\ell)}{\pi_f^{(\ell)}(e_\ell)} \\
 &= \frac{\pi_f(E')}{\pi_f(E)},
\end{align}
where $\pi_f(E)$ denotes the error probability with base error rate $p_f$ rather than $p_j$. Note that the second grouping of factors, in the first two lines, are both equal to 1 since $e'_\ell = e_\ell$. Remarkably the acceptance rate does not depend on the region rate $p_r$. Furthermore, when the flip rate is chosen to be $p_f = p_j$, then $A(E', E) = 1$, since the two ratios in Eq. \eqref{eq:acceptance} cancel. Resampling a portion of the error pattern at the same $p_j$, is clearly a valid sample from $\pi_j(E)$, so Metropolis rejections aren't necessary. In order to sample from logical failure patterns, $\pi_j(E \mid \failingset)$, rejecting correctable errors is still required.

This approach therefore offers two adjustable parameters: $p_r$ which controls the number of locations that are resampled at each step, and $p_f$ which balances the probability of applying individual errors against the overall Metropolis acceptance rate.  By varying these parameters, the resulting Metropolis-Hastings algorithm interpolates between an algorithm similar to the BV Metropolis algorithm and Monte Carlo, which resamples every location. Indeed, consider the following two regimes:

\begin{itemize}
\item \textbf{Full Resampling}: $(p_r, p_f) = (1, p_j)$: In this case, since the region rate is unity, the entire error pattern is resampled at each iteration, and individual errors are selected with their assigned physical error probabilities. Since $p_f = p_j$, we have $A(E', E) = 1$, so logical failures are always accepted.  This will generate the same set of uncorrectable patterns as standard Monte Carlo, with error pattern $E$ repeated, on average, in proportion to $\pi_i(E \mid \failingset).$
\item \textbf{Single Location Resampling}: $(p_r, p_f) = (\frac{1}{N}, p_j)$ or $(p_r, p_f) = (\frac{1}{N}, \frac{1}{2})$.  Here the resampled region typically includes a single location, and the Markov chains generated will look similar to those of the BV Metropolis algorithm.  We can push the probability of generating a new pattern into the flip rate or the acceptance probability.  
\end{itemize}

We conjecture that increasing the size of the resample region, at least for low $p_r$, decreases the decorrelation time. Imagine starting a Markov chain with $E_1$, a typical error pattern at base error rate $p_j$. Ignoring the effect of decoder rejections, $E_1$ and $E_K$ become uncorrelated after every location has been resampled. Doubling the number of locations sampled per step should approximately halve the decorrelation time. This intuition fails for larger regions since the proposed error patterns are rarely logical errors. Indeed, in the Monte Carlo regime error patterns are completely decorrelated but failures are extremely rare.  

Observations outlined in section~\ref{sec:pruning}, about fluff vs core errors, provide a good heuristic for balancing these extremes, and lead us to consider the following intermediate regime: 
\begin{itemize}
\item \textbf{Core Resampling}: $(p_r, p_f) = (1/\minfailwt, p_j)$: In this regime, the region is expected to intersect and resample the core malignant error at one location on average.  New proposals are always accepted.  
\end{itemize}
While the BV proposal algorithm selects a single location anywhere in the circuit, in this latter regime, the subregion algorithm typically selects a single critical error somewhere along the logical. This facilitates an aggressive exploration of the logical failure space, sampling error patterns with various core errors each with various arrangements of fluff errors.  At higher region rates, we would resample, and typically erase, additional errors along the core which would frequently result in a correctable error pattern. In the following, when we use the term subregion MCMC without further context, we are referring to this regime.  

\begin{figure} 
\centering
\includegraphics[width =\linewidth]{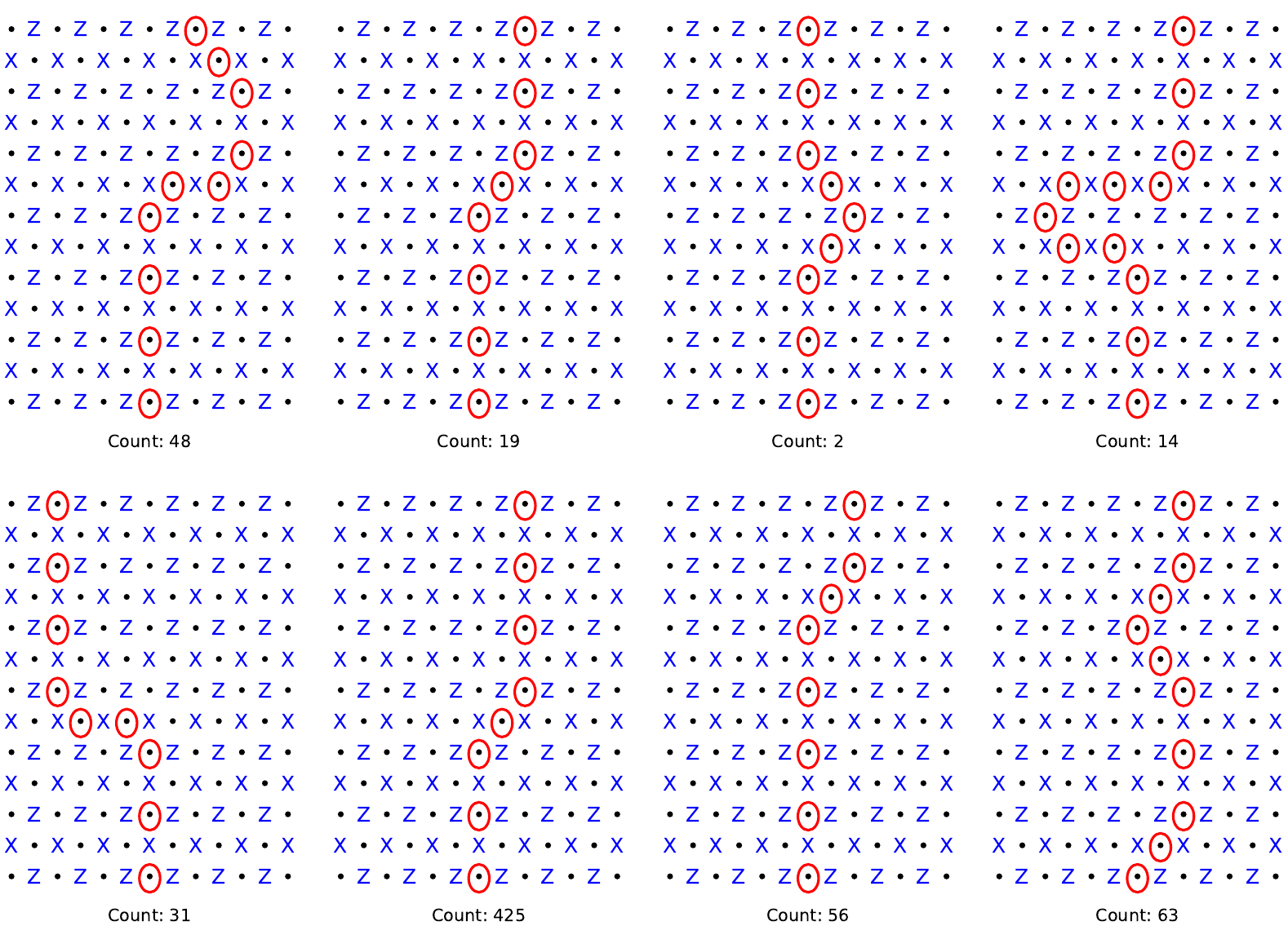}
\caption{The first eight logical failures explored by subregion MCMC, along with the number of times each pattern was included in the resulting chain, for a $d=7$ surface code at $p=1 \times 10^{-3}$.  This figure was generated by: (1) Running subregion MCMC and recording the Markov chain generated. (2) Pruning each pattern in the chain using algorithm~\ref{alg:pruning}. (3) Finally, simulating the resulting error patterns and running the decoder to determine the predicted correction.  The images shown are the qubits that contain errors after the decoder's correction was applied - the combination of the stochastically applied errors and the decoder's correction form a complete logical operator.  Repeated counts are due to either (1) sampling generated a new pattern that did not constitute a logical failure or (2) the new pattern did constitute a failure, but differed only in the fluff and not along the logical core, so the pruned patterns appear identical.  Note that in this time period, due to its high rejection rate, the Markov chain generated by BV MCMC will typically contain only repeated copies of the first logical.}
\label{fig:logicals}
\end{figure}

Figure~\ref{fig:logicals} shows an example of how the space of logical $\bar{Z}$ errors is explored.  This plot is created by first generating a Markov chain using the subregion method, then pruning the resulting failure patterns using algorithm~\ref{alg:pruning}.  This groups error patterns that share a common core but differ in their fluff.  Observe that subregion MCMC moves through the logical failure space quite efficiently, exploring a variety of dissimilar logical representatives in a small number of steps.  

Comparing the Markov graph for the BV MCMC, $\markovgraph_{\text{BV}}$, with the graph for the subregion MCMC, $\markovgraph_\text{SR}$ provides several insights. $\markovgraph_\text{BV}$ has a very sparse edge set, connecting only error patterns with hamming distance 1. In Ref. \cite{bravyi2013simulation}, the authors prove the graph is connected for surface codes, i.e., there is a sequence of weight-1 transitions connecting every logical error. $\markovgraph_\text{SR}$ however is technically a complete graph with a non-zero, albeit potentially small, probability of transition between any two error patterns, $E$ and $E'$. There is some non-zero probability that the resampling region will include every location where these patterns differ, and with some non-zero probability the resampling will produce exactly the changes required. In practice, a more relevant property is the associated mixing-time. 

\subsection{Results and Comparisons} \label{sec:results}

\begin{figure}
\centering

\begin{tikzpicture}
\begin{loglogaxis}[
  width=15cm,
  height=10.5cm,
  xlabel={Physical Error Probability},
  ylabel={Logical Error Probability},
  grid=both,
  minor grid style={gray!20},
  major grid style={gray!40},
  xtick={2e-4, 3e-4,5e-4,1e-3,2e-3},
  xticklabels={{$2\times10^{-4}$},{$3\times10^{-4}$},{$5\times10^{-4}$},{$10^{-3}$},{$2\times10^{-3}$}},
  minor x tick num=8,
  xmin=1.5e-4,
  xmax=2e-3,
  tick label style={font=\scriptsize},
  xlabel style={font=\scriptsize},
  ylabel style={font=\scriptsize},
  legend cell align=left,
 legend pos=south east,
  legend style={
    font=\tiny,      % or \footnotesize, \tiny
    fill=white,             % or e.g. fill=white,
    draw=black,      }       % or draw=black
]

% --- Surface ---
\addplot[
  only marks,
  mark=*,
  mark size=2pt,
  color=red,
  mark options={draw=red, fill=red},
  error bars/.cd,
  y dir=both,
  y explicit,
  error bar style={color=black, line width=1pt},
  error mark=|,
  error mark options={
    line width=1pt,
    mark size=6pt
  },
]
table[x=x, y=y, y error=dy] {
x           y              dy
 
1.5e-3      1.04e-4        1.04e-6       
1.21476e-3  2.36e-5        2.124e-7      
9.94357e-4  6.01e-6        5.409e-8      
7.35042e-4  7.78e-7        1.80596e-8    
5.0e-4      6.143e-8       2.15005e-9    
3.0e-4      2.42e-9        1.13256e-10   
};

% --- BS ---
\addplot[
  only marks,
  mark=triangle*,
  mark size=2.25pt,
  color=orange,
  mark options={draw=orange, fill=orange},
  error bars/.cd,
  y dir=both,
  y explicit,
  error bar style={color=black, line width=0.8pt},
  error mark=|,
  error mark options={
    line width=.75pt,
    mark size=4pt
  },
]
table[x=x, y=y, y error=dy] {
x           y              dy
6.0e-4      1.90e-4        1.90e-6       
5.05748e-4  4.57e-5        4.57e-7        
4.57482e-4  1.94e-5        4.462e-7      
4.03313e-4  6.60e-6        6.402e-8       
3.50818e-4  1.91e-6        4.775e-8       
2.78103e-4  2.69e-7        1.3181e-8      
2.0e-4      1.31e-8        1.09516e-9     
};

% Surface
\addplot[
  only marks,
  mark=o,
  mark size=2pt,
  color=blue,
  mark options={draw=blue},
]
table[x=x, y=y] {
x               y
1.500000e-03 1.040000e-04
1.457400e-03 8.548415e-05
1.415410e-03 7.015029e-05
1.374050e-03 5.656113e-05
1.333300e-03 4.622977e-05
1.293170e-03 3.735659e-05
1.253650e-03 3.015511e-05
1.214760e-03 2.426538e-05
1.176480e-03 1.932445e-05
1.138820e-03 1.544213e-05
1.101780e-03 1.227814e-05
1.065360e-03 9.765974e-06
1.029550e-03 7.768322e-06
9.943570e-04 6.156072e-06
9.597840e-04 4.884911e-06
9.258280e-04 3.809052e-06
8.924880e-04 2.976459e-06
8.597660e-04 2.295634e-06
8.276600e-04 1.773522e-06
7.961710e-04 1.365062e-06
7.652980e-04 1.033634e-06
7.350420e-04 7.838085e-07
7.054020e-04 5.911433e-07
6.763780e-04 4.443026e-07
6.479700e-04 3.338712e-07
6.201770e-04 2.497685e-07
5.930010e-04 1.873560e-07
5.664400e-04 1.390199e-07
5.404940e-04 1.020548e-07
5.151630e-04 7.405975e-08
4.904480e-04 5.419034e-08
4.663470e-04 3.902111e-08
4.428610e-04 2.796695e-08
4.199890e-04 1.999941e-08
3.977310e-04 1.412174e-08
3.760870e-04 9.822519e-09
3.550570e-04 6.768174e-09
3.346410e-04 4.627012e-09
3.148370e-04 3.109018e-09
3.000000e-04 2.269374e-09

};

% BS
\addplot[
  only marks,
  mark=o,
  mark size=2pt,
  color=violet,
  mark options={violet, fill=none},
]
table[x=x, y=y] {
x          y
6.000000e-04 1.900000e-04
5.902110e-04 1.667801e-04
5.805030e-04 1.459943e-04
5.708760e-04 1.264123e-04
5.613290e-04 1.096640e-04
5.518640e-04 9.551034e-05
5.424790e-04 8.250503e-05
5.331750e-04 7.155780e-05
5.239520e-04 6.156190e-05
5.148100e-04 5.296535e-05
5.057480e-04 4.576758e-05
4.967680e-04 3.944533e-05
4.878680e-04 3.375635e-05
4.790490e-04 2.905670e-05
4.703100e-04 2.485922e-05
4.616530e-04 2.098987e-05
4.530760e-04 1.787313e-05
4.445800e-04 1.531478e-05
4.361650e-04 1.292555e-05
4.278310e-04 1.095168e-05
4.195780e-04 9.291437e-06
4.114050e-04 7.830945e-06
4.033130e-04 6.633432e-06
3.953020e-04 5.561186e-06
3.873710e-04 4.649129e-06
3.795210e-04 3.885766e-06
3.717530e-04 3.236327e-06
3.640640e-04 2.699235e-06
3.564570e-04 2.243140e-06
3.489300e-04 1.876907e-06
3.414840e-04 1.565497e-06
3.341190e-04 1.300922e-06
3.268350e-04 1.072237e-06
3.196310e-04 8.791965e-07
3.125080e-04 7.155913e-07
3.054660e-04 5.865699e-07
2.985040e-04 4.783041e-07
2.916230e-04 3.909953e-07
2.848230e-04 3.162094e-07
2.781030e-04 2.557932e-07
2.714640e-04 2.072900e-07
2.649060e-04 1.654341e-07
2.584290e-04 1.329143e-07
2.520320e-04 1.058918e-07
2.457160e-04 8.444873e-08
2.394800e-04 6.716633e-08
2.333250e-04 5.260796e-08
2.272510e-04 4.210875e-08
2.212570e-04 3.316868e-08
2.153440e-04 2.610315e-08
2.095120e-04 2.047611e-08
2.037600e-04 1.603017e-08
2.000000e-04 1.354860e-08

};

\legend{Surface Monte Carlo, Bacon-Shor Monte Carlo, Surface MCMC, Bacon-Shor MCMC}

\end{loglogaxis}
\end{tikzpicture}

\caption{Encoded error rates for the $d=11$ surface code and the $d=5$ on $d=5$ concatenated Bacon-Shor code generated using subregion MCMC and standard Monte Carlo for comparison.  Here we seed $M=50$ chains and run until $\hat{R} \leq 1.05$.  }
\label{fig:mcmc_data}
\end{figure}
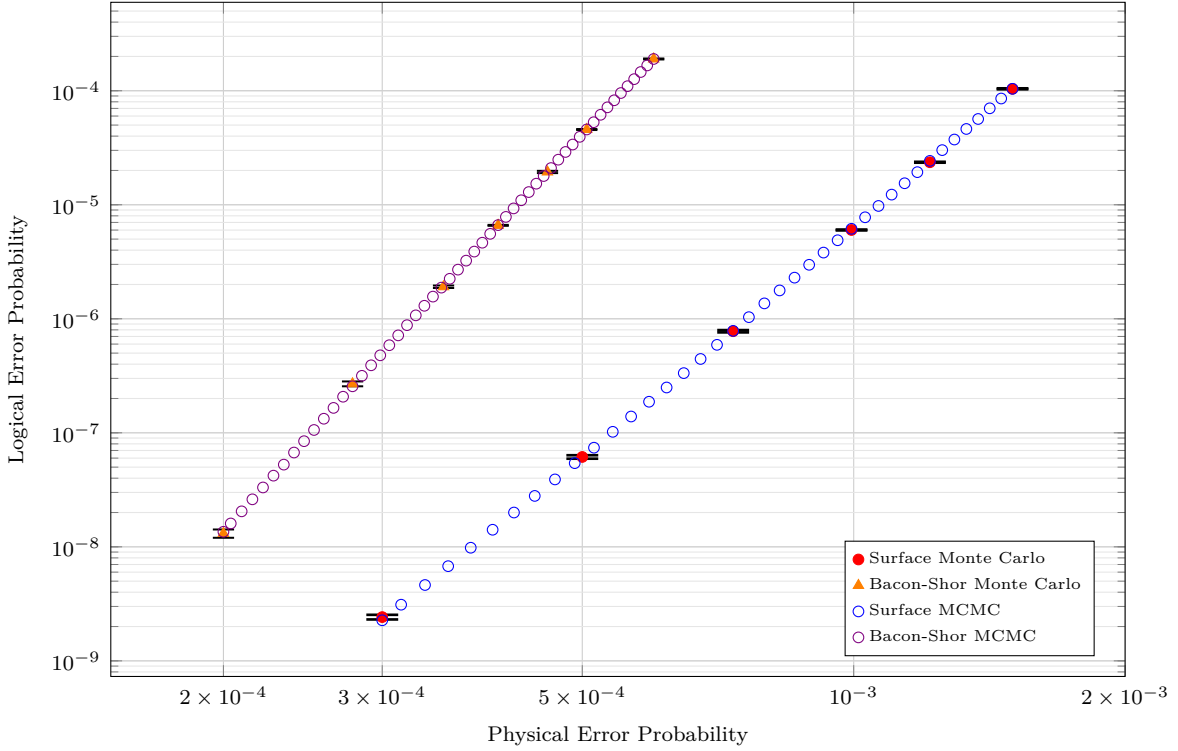

Figure~\ref{fig:mcmc_data} shows logical error rates predicted via both subregion MCMC and standard Monte Carlo for both the surface and concatenated Bacon-Shor codes.  Subregion MCMC was run using $M=50$ independent chains, each initialized at a high error rate using standard Monte Carlo, similar to Refs. \cite{mayer2025rare} and \cite{beverland2025fail}.  These $M$ chains are ideally executed on $M$ computational cores in parallel, but can be run in serial if computational resources are constrained.  In practice, we've found that MCMC and Monte Carlo match closely when the Gelman-Rubin metric reaches $\hat{R} < 1.01 - 1.05$.  

\begin{figure}
\centering

\begin{tikzpicture}
\begin{loglogaxis}[
  width=15cm,
  height=10.5cm,
  xlabel={Physical Error Probability},
  ylabel={Logical Error Probability},
  grid=both,
  minor grid style={gray!20},
  major grid style={gray!40},
  xtick={3e-4,5e-4,1e-3,2e-3},
  xticklabels={{$3\times10^{-4}$},{$5\times10^{-4}$},{$10^{-3}$},{$2\times10^{-3}$}},
  minor x tick num=8,
  xmin=2.5e-4,
  xmax=2e-3,
  tick label style={font=\scriptsize},
  xlabel style={font=\scriptsize},
  ylabel style={font=\scriptsize},
  legend cell align=left,
 legend pos=south east,
  legend style={
    font=\tiny,      % or \footnotesize, \tiny
    fill=white,             % or e.g. fill=white,
    draw=black,      }       % or draw=black
]

% --- Surface ---
\addplot[
  only marks,
  mark=*,
  mark size=2pt,
  color=red,
  mark options={draw=red, fill=red},
  error bars/.cd,
  y dir=both,
  y explicit,
  error bar style={color=black, line width=1pt},
  error mark=|,
  error mark options={
    line width=1pt,
    mark size=6pt
  },
]
table[x=x, y=y, y error=dy] {
x           y              dy
 
1.5e-3      1.04e-4        1.04e-6       
1.21476e-3  2.36e-5        2.124e-7      
9.94357e-4  6.01e-6        5.409e-8      
7.35042e-4  7.78e-7        1.80596e-8    
5.0e-4      6.143e-8       2.15005e-9    
3.0e-4      2.42e-9        1.13256e-10   
};

% Downward
\addplot[
  only marks,
  mark=o,
  mark size=2pt,
  color=blue,
  mark options={draw=blue},
]
table[x=x, y=y] {
x               y
1.500000e-03 1.040000e-04
1.457400e-03 8.465818e-05
1.415410e-03 6.886599e-05
1.374050e-03 5.634314e-05
1.333300e-03 4.542689e-05
1.293170e-03 3.630349e-05
1.253650e-03 2.894975e-05
1.214760e-03 2.361174e-05
1.176480e-03 1.876649e-05
1.138820e-03 1.483269e-05
1.101780e-03 1.181887e-05
1.065360e-03 9.339304e-06
1.029550e-03 7.389855e-06
9.943570e-04 5.767861e-06
9.597840e-04 4.541534e-06
9.258280e-04 3.547461e-06
8.924880e-04 2.769302e-06
8.597660e-04 2.154339e-06
8.276600e-04 1.659767e-06
7.961710e-04 1.273719e-06
7.652980e-04 9.749438e-07
7.350420e-04 7.376044e-07
7.054020e-04 5.609001e-07
6.763780e-04 4.238572e-07
6.479700e-04 3.147477e-07
6.201770e-04 2.293418e-07
5.930010e-04 1.680494e-07
5.664400e-04 1.217518e-07
5.404940e-04 8.939518e-08
5.151630e-04 6.606746e-08
4.904480e-04 4.726280e-08
4.663470e-04 3.358930e-08
4.428610e-04 2.326334e-08
4.199890e-04 1.629992e-08
3.977310e-04 1.118582e-08
3.760870e-04 7.650479e-09
3.550570e-04 5.232906e-09
3.346410e-04 3.555770e-09
3.148370e-04 2.363234e-09
3.000000e-04 1.718371e-09
};

% Upward
\addplot[
  only marks,
  mark=o,
  mark size=2pt,
  color=green!50!black,
  mark options={draw=green!50!black, fill=none},
]
table[x=x, y=y] {
x          y
1.500000e-03 1.040000e-04
1.457400e-03 8.551868e-05
1.415410e-03 6.894004e-05
1.374050e-03 5.709174e-05
1.333300e-03 4.686230e-05
1.293170e-03 3.815937e-05
1.253650e-03 3.086294e-05
1.214760e-03 2.555956e-05
1.176480e-03 2.061207e-05
1.138820e-03 1.641838e-05
1.101780e-03 1.315142e-05
1.065360e-03 1.055850e-05
1.029550e-03 8.337659e-06
9.943570e-04 6.671215e-06
9.597840e-04 5.276939e-06
9.258280e-04 4.160520e-06
8.924880e-04 3.300107e-06
8.597660e-04 2.591358e-06
8.276600e-04 2.005401e-06
7.961710e-04 1.567654e-06
7.652980e-04 1.194180e-06
7.350420e-04 9.253348e-07
7.054020e-04 7.107994e-07
6.763780e-04 5.443994e-07
6.479700e-04 4.123610e-07
6.201770e-04 3.110962e-07
5.930010e-04 2.334353e-07
5.664400e-04 1.729239e-07
5.404940e-04 1.284962e-07
5.151630e-04 9.481150e-08
4.904480e-04 6.989091e-08
4.663470e-04 5.013715e-08
4.428610e-04 3.642028e-08
4.199890e-04 2.542207e-08
3.977310e-04 1.767834e-08
3.760870e-04 1.235354e-08
3.550570e-04 8.500419e-09
3.346410e-04 5.802087e-09
3.148370e-04 3.881197e-09
3.000000e-04 2.823985e-09
};

\legend{Monte Carlo, Downward Splitting, Upward Splitting}

\end{loglogaxis}
\end{tikzpicture}
\caption{Encoded error rates for the $d=11$ surface code, generated using subregion MCMC with $\hat{R}=1.2$.  The blue points are generated in the usual way, via downward splitting, while the green points are generated via upward splitting.  Notice that since the chains were not given enough time to burn-in, the ratios predicted by upward and downward splitting diverge. Here, unlike our other simulations, we do not throw away the first half of each chain so this effect is more obvious. }
\label{fig:upward}
\end{figure}
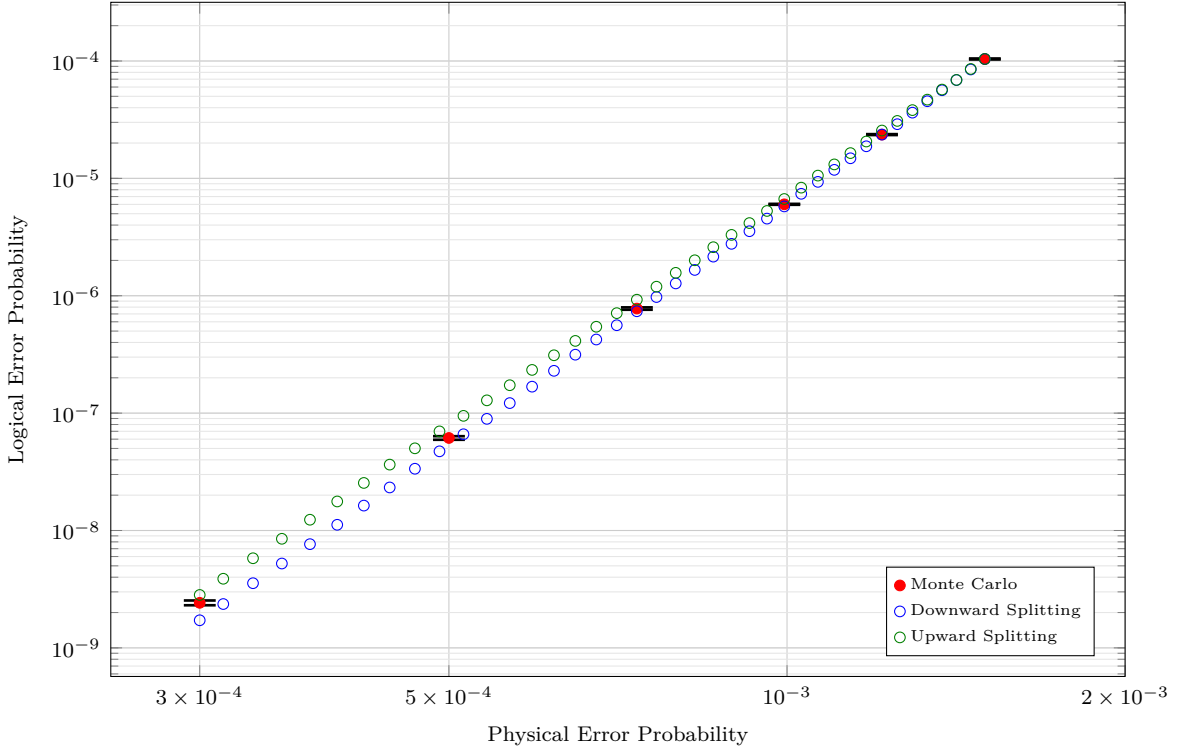

If chains are especially short, they may not make it past the ``burn-in'' period,  where they still reflect the distribution from the previous step. During this time, the failure patterns generated may be more consistent with the higher probability $p_{j-1}$, rather than $p_j$.  This means that the weight of the generated error chains will often be higher than expected, and consequently, the computed ratio can be biased low.  This effect is reversed during ``upward splitting,'' where MCMC begins with a low error rate.   This is illustrated in Fig.~\ref{fig:upward}, where chains are terminated with a high $\hat{R} \leq 1.2$, and therefore have not had sufficient time to converge.  These effects are mitigated by throwing away the first half of the samples in each chain.  We've found this practice, combined with a lower $\hat{R}$ provide a better match to Monte Carlo generated logical error rates.

In order to compare the runtime of various MCMC methods, we count the number of circuit simulations required for convergence. Convergence is  measured by ensuring the Gelman-Rubin statistic is below a fixed value, $\hat{R}$, when using a fixed number, $M$, of Markov chains. The work required to sample from the proposal distribution is negligible compared to circuit simulation and decode. Given the connection between $\hat{R}$ and effective sample size \cite{vats2021revisiting}, this metric is equivalent to the number of circuit simulations required to produce some number of effective samples.  Since BV MCMC was originally described in terms of bit or phase flips, here we switch to a bit flip error model, where an $X$ error is applied after each gate to each qubit with equal probability $p_j$.  

\begin{figure}
\centering
\begin{subfigure}{0.49\textwidth}
\centering
\begin{tikzpicture}
\begin{axis}[
  width=\linewidth,
  height=7cm,
  xlabel={Distance},
  ylabel={Number of Decodes},
  tick label style={font=\scriptsize},
  xlabel style={font=\scriptsize},
  ylabel style={font=\scriptsize, yshift = -14pt},
  grid=both,
  xtick={7,9,11,13,15,17},
  xticklabels={{$7$},{$9$},{$11$},{$13$},{$15$},{$17$}},
  ytick={5e5,1e6,1.5e6,2.0e6,2.5e6},
  yticklabels={{$.5$},{$1$},{$1.5$},{$2.0$},{$2.5$}},
  scaled y ticks=true,
  minor grid style={gray!20},
  major grid style={gray!40},
  legend pos=north west,
  legend style={font=\scriptsize},
  legend cell align=left,
]

% timing
\addplot[
  only marks,
  mark=*,
  mark size=1.5pt,
  color=red,
  mark options={draw=red, fill=red},
]
coordinates {
  (7,  236757)
  (9,  529862)
  (11, 873682)
  (13, 1638970)
  (15, 1981019)
  (17, 2756927)
};

% timingRegion
\addplot[
  only marks,
  mark=*,
  mark size=1.5pt,
  color=blue,
  mark options={draw=blue, fill=blue},
]
coordinates {
  (7,  129832)
  (9,  174997)
  (11, 190000)
  (13, 280000)
  (15, 310000)
  (17, 305000)
};

% timingBS
\addplot[
  only marks,
  mark=triangle*,
  mark size=1.75pt,
  color=orange,
  mark options={draw=orange, fill=orange},
]
coordinates {
  (7,  242902)
  (17, 1533138)
};

% timingBSRegion
\addplot[
  only marks,
  mark=triangle*,
  mark size=1.75pt,
  color=violet,
  mark options={draw=violet, fill=violet},
]
coordinates {
  (7,  84999)
  (17, 145000)
};

\legend{Surface BV, Surface subregion, Bacon-Shor BV, Bacon-Shor subregion}

\end{axis}
\end{tikzpicture}
\caption{}
\end{subfigure}
\hfill
\begin{subfigure}{0.49\textwidth}
\centering
\begin{tikzpicture}
\begin{axis}[
  width=\linewidth,
  height=7cm,
  xlabel={Physical Error Probability},
  ylabel={Number of Decodes},
  grid=both,
  tick label style={font=\scriptsize},
  xlabel style={font=\scriptsize},
  ylabel style={font=\scriptsize, yshift = -16pt},
  ytick={1e5,2e5,3e5,4e5,5e5,6e5},
  yticklabels={{$1$},{$2$},{$3$},{$4$},{$5$},{$6$}},
  xtick={.0005,.001,.0015},
  xticklabels={{$.0005$},{$.001$},{$.0015$}},
  minor grid style={gray!20},
  major grid style={gray!40},
  legend pos=south east,
  legend style={font=\scriptsize},
  legend cell align=left,
  scaled y ticks=true,
  scaled x ticks=false,
  y tick label style={/pgf/number format/fixed},
  x dir=reverse,             % <-- flip x-axis
]

% y1: blue filled circles
\addplot[
  only marks,
  mark=*,
  mark size=1.5pt,
  color=red,
  mark options={draw=red, fill=red},
]
coordinates {
  (0.0015,      287644)
  (0.00134437,  586038)
  (0.00119748,  500865)
  (0.00105931,  540232)
  (0.000929853, 532383)
  (0.000809081, 560052)
  (0.000696975, 430538)
  (0.000593509, 511648)
  (0.000498652, 466644)
  (0.000412371, 278523)
  (0.000334624, 333706)
};

% y2: red hollow circles
\addplot[
  only marks,
  mark=*,
  mark size=1.5pt,
  color=blue,
  mark options={draw=blue, fill=blue},
]
coordinates {
  (0.0015,      60000)
  (0.00134437,  130000)
  (0.00119748,  160000)
  (0.00105931,  200000)
  (0.000929853, 240000)
  (0.000809081, 169997)
  (0.000696975, 149988)
  (0.000593509, 149696)
  (0.000498652, 159889)
  (0.000412371, 179637)
  (0.000334624, 218881)
};

\legend{Surface BV, Surface subregion}

\end{axis}
\end{tikzpicture}

\caption{}
\end{subfigure}
\caption{(a) A comparison of the runtime of subregion and BV MCMC for surface codes of distances 7,9,11,13,15, and 17 at $p = 1.5 \times 10^{-3}$ and for concatenated Bacon-Shor codes at distances 3-on-3 and 5-on-5 at  $p = 6 \times 10^{-4}$. The count at a given distance is the number of decodes needed to compute the first MCMC ratio, i.e., the number of decodes required for the first two chains.  Blue dots correspond to subregion MCMC applied to the surface code, red dots to BV MCMC applied to the surface code, purple triangles to subregion MCMC applied to the Bacon-Shor code, and orange triangles to BV MCMC applied to the Bacon-Shor code.  The concatenated Bacon-Shor code points are plotted at their effective distance, $d = 2 \minfailwt + 1$. (b) The number of decodes needed to compute the Markov chains for a $d = 11$ surface code at each $p_j$ as we step down from $p_j = 1.5 \times 10^{-3}$ to $p_j =  3 \times 10^{-4}$.  For both (a) and (b), we use a bit flip error model, run $M = 50$ chains and terminate when $\hat{R} \le 1.05$, after discarding the first half of each chain. This corresponds to about 9.75 independent samples per chain, or a total of about 487.}
\label{fig:timing}
\end{figure}
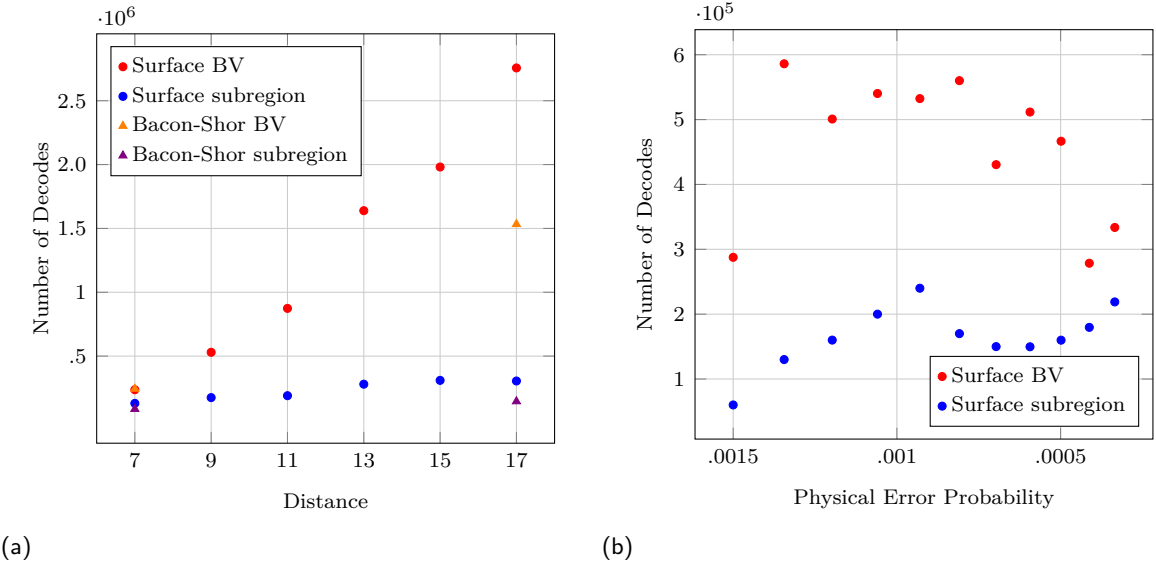

Fig.~\ref{fig:timing} demonstrates a speedup in subregion MCMC of between approximately 2X and 10X over BV MCMC, with a growing improvement as a function of increasing distance.  We conjecture that the mixing-time of subregion MCMC should be similarly faster; resampling about $N/d$ locations in each step allows the error pattern to more easily jump from one logical class to another.  In the low $p$ regime, it may be beneficial to adaptively adjust $(p_r, p_f)$.  Indeed, while not illustrated here, the performance of both approaches can suffer as the number of applied physical errors becomes sparse, i.e., when the expected number of errors is less than $t$.  We leave this for future work.  

\section{Conclusions and Future Work}

Here we have described a simple and fast algorithm for finding minimum weight failure patterns in a quantum error correction circuit. Algorithm~\ref{alg:pruning} is useful for debugging both circuits and decoding algorithms. The algorithm is motivated by, and demonstrates the utility of, the fluff vs core concept. 

In addition, we have introduced subregion MCMC, which uses a new family of proposal distributions. Subregion MCMC is extremely fast in practice.  The MCMC data points in Fig.~\ref{fig:mcmc_data} can be generated in hours on a desktop PC, while the matching Monte Carlo points require many days and/or a high performance cluster.  Moreover, as illustrated above, subregion MCMC significantly outperforms BV MCMC, in terms of the number of simulations required for convergence. This makes the simulation of many codes at utility-scale error rates entirely feasible, even without significant computational hardware. 

A potentially beneficial task for future work would be to incorporate a reliable estimate for statistical error. A commonly used metric is the Monte Carlo Standard Error \cite{geyer2011mcmc}, which properly accounts for correlations among the generated samples. Typical approaches include estimating the effective sample size (ESS), or equivalently, the average decorrelation time, from the chain data. It's important to note this error does not quite furnish error bars for the estimated ratios and/or logical errors rates unless the chains are fully mixed. False convergence might effectively introduce a bias, in which case the statistical error is only one source of error. 

Subregion MCMC demonstrates the value of investigating new proposal algorithms. By parameterizing our algorithm, we have shown that there is a continuum of sampling strategies between Monte Carlo and MCMC.  With heuristically chosen parameters $(p_r, p_f) = (1/\minfailwt, p_j)$, motivated using the fluff vs. core concept, this method already offers an order of magnitude improvement in the number of simulations required for a given effective sample size. Optimizing these parameters could improve performance significantly, especially when the expected number of errors drops below $\minfailwt$. The Metropolis-Hastings framework provides a recipe for creating entirely new MCMC methods. All that's required is an algorithm for generating new error patterns and the ratio of the associated proposal distributions. 

\section*{Acknowledgments} 
We would like to thank Jonas Anderson and Bryan Eastin for review and many useful discussions.

\bibliographystyle{plain}
\bibliography{mcmc_references}

@article{lee2021even,
  title={Even more efficient quantum computations of chemistry through tensor hypercontraction},
  author={Lee, Joonho and Berry, Dominic W and Gidney, Craig and Huggins, William J and McClean, Jarrod R and Wiebe, Nathan and Babbush, Ryan},
  journal={PRX quantum},
  volume={2},
  number={3},
  pages={030305},
  year={2021},
  publisher={APS}
}

@article{gidney2025factor,
  title={How to factor 2048 bit RSA integers with less than a million noisy qubits},
  author={Gidney, Craig},
  journal={arXiv preprint arXiv:2505.15917},
  year={2025}
}

@article{beverland2025fail,
  title={Fail fast: techniques to probe rare events in quantum error correction},
  author={Beverland, Michael E and Carroll, Malcolm and Cross, Andrew W and Yoder, Theodore J},
  journal={arXiv preprint arXiv:2511.15177},
  year={2025}
}

@article{bennett1976efficient,
  title={Efficient estimation of free energy differences from {Monte} {Carlo} data},
  author={Bennett, Charles H},
  journal={Journal of Computational Physics},
  volume={22},
  number={2},
  pages={245--268},
  year={1976},
  publisher={Elsevier}
}

@article{bravyi2013simulation,
  title={Simulation of rare events in quantum error correction},
  author={Bravyi, Sergey and Vargo, Alexander},
  journal={Physical Review A—Atomic, Molecular, and Optical Physics},
  volume={88},
  number={6},
  pages={062308},
  year={2013},
  publisher={APS}
}

@incollection{gelman2011inference,
  author    = {Gelman, Andrew and Shirley, Kenneth},
  title     = {Inference from Simulations and Monitoring Convergence},
  booktitle = {Handbook of Markov Chain Monte Carlo},
  editor    = {Brooks, Steve and Gelman, Andrew and Jones, Galin and Meng, Xiao-Li},
  publisher = {Chapman and Hall/CRC},
  year      = {2011}, 
  chapter   = {6},
  pages     = {163--174},
  doi       = {10.1201/b10905-7},
  isbn      = {978-1-4200-7941-8}
}

@incollection{geyer2011mcmc,
  author    = {Geyer, Charles},
  title     = {Introduction to {Markov} Chain {Monte Carlo}},
  booktitle = {Handbook of Markov Chain Monte Carlo},
  editor    = {Brooks, Steve and Gelman, Andrew and Jones, Galin and Meng, Xiao-Li},
  publisher = {Chapman and Hall/CRC},
  year      = {2011}, 
  chapter   = {1},
  pages     = {3--48},
  doi       = {10.1201/b10905-7},
  isbn      = {978-1-4200-7941-8}
}

@article{mayer2025rare,
  title={Rare Event Simulation of Quantum Error-Correcting Circuits},
  author={Mayer, Carolyn and Ganti, Anand and Onunkwo, Uzoma and Metodi, Tzvetan and Anker, Benjamin and Skryzalin, Jacek},
  journal={arXiv preprint arXiv:2509.13678},
  year={2025}
}

@article{kolmogorov2009blossom,
  title={Blossom V: a new implementation of a minimum cost perfect matching algorithm},
  author={Kolmogorov, Vladimir},
  journal={Mathematical Programming Computation},
  volume={1},
  number={1},
  pages={43--67},
  year={2009},
  publisher={Springer}
}

@article{heussen2024dynamical,
  title={Dynamical subset sampling of quantum error-correcting protocols},
  author={Heu{\ss}en, Sascha and Winter, Don and Rispler, Manuel and M{\"u}ller, Markus},
  journal={Physical Review Research},
  volume={6},
  number={1},
  pages={013177},
  year={2024},
  publisher={APS}
}

@article{vats2021revisiting,
  title={Revisiting the {Gelman-Rubin} Diagnostic},
  author={Dootika Vats and Christina Knudson},
  volume = {36},
  journal = {Statistical Science},
  number = {4},
  publisher = {Institute of Mathematical Statistics},
  pages =  {518 -- 529},
  year = {2021},
  doi = {10.1214/20-STS812}
}

@article{dennis2002topological,
  title={Topological quantum memory},
  author={Dennis, Eric and Kitaev, Alexei and Landahl, Andrew and Preskill, John},
  journal={Journal of Mathematical Physics},
  volume={43},
  number={9},
  pages={4452--4505},
  year={2002},
  publisher={American Institute of Physics}
}

@article{bacon2006operator,
  title={Operator quantum error-correcting subsystems for self-correcting quantum memories},
  author={Bacon, Dave},
  journal={Physical Review A—Atomic, Molecular, and Optical Physics},
  volume={73},
  number={1},
  pages={012340},
  year={2006},
  publisher={APS}
}

@article{aliferis2005quantum,
  title={Quantum accuracy threshold for concatenated distance-3 codes},
  author={Aliferis, Panos and Gottesman, Daniel and Preskill, John},
  journal={Quantum Information \& Computation},
  volume={6},
  number={2},
  pages={97--165},
  year={2006},
  publisher={Rinton Press, Incorporated Paramus, NJ}
}

@article{metropolis1953equation,
title={Equation of state calculations by fast computing machines},
author={Metropolis, Nicholas and Rosenbluth, Arianna W and Rosenbluth, Marshall N and Teller, Augusta H and Teller, Edward},
journal={The journal of chemical physics},
volume={21},
number={6},
pages={1087--1092},
year={1953},
publisher={American Institute of Physics}
}

@article{hastings1970monte,
title={{Monte} {Carlo} sampling methods using Markov chains and their applications},
author={Hastings, W Keith},
year={1970},
publisher={Oxford University Press}
}

@article{geman1984stochastic,
title={Stochastic relaxation, {Gibbs} distributions, and the {Bayesian} restoration of images},
author={Geman, Stuard and Geman, Donald},
journal={IEEE Transactions on pattern analysis and machine intelligence},
number={6},
pages={721--741},
year={1984},
publisher={IEEE}
}

@article{neal2003slice,
title={Slice sampling},
author={Neal, Radford M},
journal={The Annals of Statistics},
volume={31},
number={3},
pages={705--767},
year={2003},
publisher={Institute of Mathematical Statistics}
}

@article{duane1987hybrid,
title={Hybrid {Monte} {Carlo}},
author={Duane, Simon and Kennedy, Anthony D and Pendleton, Brian J and Roweth, Duncan},
journal={Physics letters B},
volume={195},
number={2},
pages={216--222},
year={1987},
publisher={Elsevier}
}

\end{document}